\documentstyle[12pt,epsf]{article}
\newcommand{\beq}{\begin{equation}}
\newcommand{\eeq}{\end{equation}}
\newcommand{\beqa}{\begin{eqnarray}}
\newcommand{\eeqa}{\end{eqnarray}}

\begin{document}


\hfill LPT 96--14

\hfill TK 96 20

\hfill hep-ph/9607nnn

\bigskip\bigskip

\begin{center}

{{\large\bf On the analysis of the pion--nucleon $\sigma$--term:

\vskip 0.2 cm

The size of the remainder at the Cheng--Dashen point}}

\end{center}

\vspace{.5in}

\begin{center}
{\large V. Bernard$^a$, N. Kaiser$^b$, 
Ulf-G. Mei{\ss}ner$^c$}\footnote{Address after Oct. 1, 1996: FZ
J\"ulich, IKP (Theorie), D-52425 J\"ulich, Germany}

\bigskip

\bigskip

$^a$Universit\'e Louis Pasteur, Laboratoire de Physique Th\'eorique\\
BP 28, F-67037 Strasbourg Cedex 2, France\\ 
{\it email: bernard@crnhp4.in2p3.fr}\\
\vspace{0.3cm}
$^b$Technische Universit\"at M\"unchen, Physik Department T39\\
James-Franck-Stra\ss e, D-85747 Garching, Germany\\ 
{\it email: nkaiser@physik.tu-muenchen.de}\\
\vspace{0.3cm}
$^c$Universit\"at Bonn, Institut f{\"u}r Theoretische Kernphysik\\
Nussallee 14-16, D-53115 Bonn, Germany\\
{\it email: meissner@itkp.uni-bonn.de}\\
\end{center}

\vspace{0.5in}

\thispagestyle{empty}

\begin{abstract}
\noindent We calculate the one--loop contributions of order $M_\pi^4$ to the
difference $\Delta_R$ between the on--shell pion--nucleon scattering amplitude
$\bar{D}^+(0,2M_\pi^2)$ at the Cheng--Dashen point $\nu =0$, $t=2M_\pi^2$ and
the scalar form factor $\sigma (2M_\pi^2)$ in the framework of heavy baryon 
chiral perturbation theory. We proof that to this order $\Delta_R$ contains
$no$ chiral logarithms and therefore it vanishes simply as $M_\pi^4$ in the
chiral limit. Numerically, we find as an upper limit $\Delta_R \simeq 2\,$MeV.
\end{abstract}

\vfill

\today 

\newpage

\noindent {\bf 1.} 
Pion--nucleon scattering data allow to extract information on the size
of the pion--nucleon  $\sigma$--term, $\sigma (0)$, which measures 
the explicit chiral symmetry breaking in QCD due to the up-- and
down--quark masses. A venerable (current algebra) low--energy theorem  due to 
Brown, Pardee and Peccei \cite{bpp} relates $\sigma (0)$ to the isoscalar
$\pi N$ scattering amplitude (with the pseudovector Born term subtracted) via 
\begin{equation} 
F_\pi^2 \, \bar D^+(0,2M_\pi^2) - \sigma(2M_\pi^2) = 
F_\pi^2 \, \bar D^+(0,2M_\pi^2) -\Delta_\sigma - \sigma (0) = \Delta_R
= M_\pi^4 \, C_R 
\end{equation}
with $F_\pi \, (M_\pi)$ the pion decay constant (mass) and 
$\Delta_\sigma = \sigma(2M_\pi^2) - \sigma(0)$. 
The crucial statement of the low--energy theorem is that the 
remainder $\Delta_R$ grows quadratically with the light quark mass. 
The on--shell
$\pi N$ scattering amplitude $\pi^a(q)+N(p)\to \pi^b(q')+N(p')$, with 
$s=(p+q)^2$ and $t= (p-p')^2$ the conventional Mandelstam variables,
is defined in the standard fashion,
\begin{equation}
 T^{ba}_{\pi N}= \bar u (p')\bigl\{ \delta^{ba} \bigl[ A^+(s,t) +
q\!\!\!/ \, B^+(s,t)\bigr] + i\epsilon^{bac}\tau^c \bigl[ A^-(s,t)+
 q\!\!\!/  \,
B^-(s,t) \bigr] \bigr\} u(p)  \,\, .
\end{equation}
We also introduce the variable $\nu = (s-m^2+t/2-M_\pi^2)/2m$, with $m$ the
nucleon mass. At the Cheng--Dashen point $\nu = 0,\, s=m^2,\, t = 2M_\pi^2$
\cite{cd}, 
\begin{equation} 
\bar D^+(0,2M_\pi^2) = A^+(m^2,2M_\pi^2) - {g_{\pi N}^2 \over m} \,\,
\, ,
\end{equation}
with the last term due to the subtraction of the pseudovector tree amplitude.
Furthermore, the nucleon scalar form factor $\sigma(t)$ is given by the matrix
element  
\begin{equation}
 <N(p') |\hat m(\bar u u+\bar dd)|N(p)> = \bar u(p')\, \sigma(t)\, u(p) 
\,\, ,\,\, t = (p-p')^2 \,\, \, .
\end{equation}
Although the Cheng--Dashen point is not in the
physical region of the $\pi N$ scattering process, it lies well
within the Lehmann ellipse and thus $\bar D^+(0,2M_\pi^2)$ can 
 be obtained by analytic
continuation, i.e. using dispersion relations. The most recent
analysis leads to $F_\pi^2 \,\bar D^+(0,2M_\pi^2) = 60 \, (62) \, $MeV for
two sets of low energy $\pi N$ data \cite{gls}. For a discussion of the 
uncertainties (typically $\pm 8$ MeV) and previous determinations, we refer to
\cite{gls}.  The leading non--analytic contribution to the
scalar form factor difference, $\Delta_\sigma$, is $3g_{\pi N}^2
M_\pi^3 /(64 \pi m^2) $ and gives about 8 MeV 
\cite{pp} \cite{gss}. Evaluating the same one--loop
diagram with an intermediate $\Delta(1232)$ isobar adds another 7 MeV 
\cite{bkmz}. A  detailed dispersive analysis \cite{gls}, with $\pi \pi$ and
$\pi N$  information consistent with chiral symmetry, yields $\Delta_\sigma =
15.2 \pm 0.4 \,$MeV. 

The remainder $\Delta_R$ is not fixed by chiral
symmetry. It has to be known, however, to extract information on the
$\sigma$--term, i.e. $\sigma(0)$, and from it the strangeness content of the
proton (the matrix element $<p|\, \bar s s \,|p>$). 
Let us briefly summarize what
is known about the size of $\Delta_R$. Brown et al.\cite{bpp} estimated the
remainder from tree level resonance excitation, with most of its contribution
coming from the $\Delta(1232)$, of the order of 0.6 MeV. Furthermore,
it was shown that spin--$\frac{1}{2}^\pm$ resonances do not contribute to the 
isoscalar $\pi N$ amplitude at the Cheng--Dashen point while the higher
spin--$\frac{3}{2}^-$ resonance contributions are suppressed by two orders of
magnitude compared to the $\Delta(1232)$ \cite{bpp}. Gasser et al. \cite{gss}
performed a complete one--loop calculation of the $\pi N$ scattering amplitude
in relativistic nucleon chiral perturbation theory to order $q^3$ and found
$\Delta_R^{\rm (GSS)} = 0.35\,$MeV, or truncated at order $M_\pi^4$, 
\begin{equation} 
\Delta_R^{\rm (GSS)} = { g_A^4 \, M_\pi^4 \over 32 \pi^2 m F_\pi^2
}  = 0.46 \,  {\rm MeV} \,\,\, . 
\label{GSS}
\end{equation} 
Therefore, the conjecture of Pagels and Pardee \cite{pp} that
$\Delta_R$ contains potentially large logarithms of the form $M_\pi^4
\ln M_\pi$ could not be verified (to order $q^3$). 
However, there could still be
large logarithms at one--loop in diagrams which have exactly one
insertion from the dimension two effective pion--nucleon 
Lagrangian ${\cal L}_{\pi N}^{(2)}$.
Such a large effect at subleading  order has already been observed in
the calculation of the magnetic polarizability of the proton, where at
order $q^4$ the loop graphs generate a $\ln M_\pi$ term with a large
coefficient which cancels most of the big contribution from the
$\Delta(1232)$ encoded in the low--energy constant of the pertinent contact
term from ${\cal L}_{\pi N}^{(4)}$. Furthermore, some of the
coefficients of ${\cal L}_{\pi N}^{(2)}$ are considerably larger than
their natural size $1/2m \simeq 0.5\,$GeV$^{-1}$, see e.g. the 
review~\cite{bkmr}. It  therefore appears mandatory to perform a
complete ${\cal O}(q^4)$ calculation to see whether such logarithms are present
and to find a more accurate bound on the size of the remainder $\Delta_R$.
 
\medskip

\noindent {\bf 2.} The tool to systematically calculate {\it all}
corrections to a given order is chiral perturbation
theory (CHPT). It amounts to a systematic expansion around the chiral limit
in terms of two small parameters related to the quark masses and the
external momenta. To have a consistent power counting in the presence 
of baryons, the latter have to be treated as very heavy (static) sources, i.e. 
non--relativistically. We follow here the systematic SU(2) approach developed 
in Ref.\cite{bkkm}. In the framework of heavy baryon CHPT and to order $q^4$,
we have to consider pion loop diagrams with at most one insertion from
the dimension two pion--nucleon Lagrangian 
and local contact terms from ${\cal L}_{\pi N}^{(4)}$ accompanied 
by a priori unknown coefficients, the so--called
low-energy constants (LECs). These we are estimating by resonance exchange
since not enough precise data exist yet to pin them all down. However,
previous calculations have already shown that this approach of
treating the LECs is fairly accurate as long as no big cancellations
appear (for details, see \cite{bkmr}). 

Consider first the possible contact term contributions. Tree--level
$\Delta(1232)$-exchange is independent of the off-shell parameter $Z$ entering
the $\pi N\Delta$-vertex. Using the empirically well satisfied 
large $N_c$  
coupling constant relation $g_{\pi N\Delta } = 3 g_{\pi N}/\sqrt2$ and the
Goldberger--Treiman relation $g_A = g_{\pi N}F_\pi/m$, we find 
\begin{equation} 
\Delta_R^{(\Delta)} = { g_A^2 \, M_\pi^4 \over 4
m_\Delta(m_\Delta^2-m^2) }\biggl(2 + {m\over m_\Delta} \biggr)= 0.58 \, {\rm
MeV} 
\end{equation}
in good agreement with the estimate of Brown et al.\cite{bpp}. Furthermore,
$N^*(J^P=\frac{1}{2}^\pm)$ nucleon resonance exchange gives
\begin{equation} 
\Delta_R^{(N^*)} = 0 \, \, , 
\end{equation} 
verifying the general argument given in \cite{bpp} within the chiral effective 
field theory approach. Here, by chiral symmetry requirements the pion coupling
is of vector/axial-vector type for the parity odd/even spin-$\frac{1}{2}$
resonances. The situation concerning
the scalar--isoscalar meson exchange  is somewhat more complex.
Using the lowest order effective Lagrangians consistent with chiral
symmetry \cite{reso1} 
\beq
{\cal L}_{NS} = g_S \, S \, \bar{\Psi} \Psi\,\, , \quad
{\cal L}_{S\pi} = S \, [ c_m \, {\rm Tr} \chi_+ + c_d \, {\rm Tr}
(u_\mu u^\mu)]
\eeq
in the conventional notation ($S$ denotes the scalar and $\Psi$ the
nucleon field), we find
\beq
 \Delta_R^{(S,2)} = 0 \, \, .
\eeq
While the term $\sim c_m$ cancels in $\Delta_R$ 
to all orders if the scalar meson propagator is chirally expanded, the one
$\sim c_d$ is proportional to $q \cdot q'= M_\pi^2-t/2$ and thus vanishes at
the Cheng--Dashen point. Only a four--derivative scalar meson--pion coupling 
\beq
{\cal L}_{S\pi} = c_{4d} \, S \, \,  {\rm Tr}(D\cdot u)^2 
\eeq
would make a non--vanishing contribution to $\Delta_R$. 
Since from the phenomenological side essentially nothing is known 
about the strength of such a  vertex we can only give an estimate
based on dimensional arguments. The low--energy constant $c_3$ 
(defined in Eq.(\ref{LpiN2}) below) has
been determined from low-energy $\pi N$ data, $c_3 \simeq 
-4\,$GeV$^{-1}$ \cite{bkmpipin} and its value can be understood
from combined $\Delta$ and scalar--isoscalar exchange, the latter
contributing at most $c_3^{(S)} = -1\,$GeV$^{-1}$
\cite{bkmr}. Furthermore, we have  $c_3^{(S)} = 2 g_S c_d / M_S^2$ and
thus the four-derivative--scalar contribution to $\Delta_R$ takes the form
\beq
\Delta_R^{(S,4)} = 4 M_\pi^4 \, \frac{ c_{4d} \, g_S }{ M_S^2} 
=  2 M_\pi^4 \, \biggl|\frac{c_{4d}}{c_d} \, c_3^{(S)} \biggr| \,\, .
\eeq
Notice that the sign of $c_{4d}$ is not fixed, we have chosen it to
give a positive contribution to $\Delta_R$ and thus we can obtain an
upper bound on the remainder. 
Assuming now that each derivative $D_\mu$ is suppressed by $1/ 4\pi
F_\pi$, i.e. the typical scale of chiral symmetry breaking, we get
$ | c_{4d}/c_d |= 0.73\,$GeV$^{-2}$. This  gives $\Delta_R^{(S,4)} 
= 0.55\,$MeV. Allowing for a factor of two uncertainty, we arrive at
\beq
\Delta_R^{(S)} \simeq 1.1 \, {\rm MeV} \,\,\, .
\eeq

We now turn to the calculation of the one-loop graphs with exactly one
insertion from ${\cal L}_{\pi N}^{(2)}$. The dimension two
chiral $\pi N$ Lagrangian has the form \cite{bkmr,bkkm}
\begin{equation} 
{\cal L}^{(2)}_{\pi N} = \bar N \biggl\{ c_1 \,{\rm Tr}\,\chi_+
+ c_2 \,(v\cdot u)^2 + c_3\, u_\mu u^\mu + c_4\, [S^\mu,S^\nu] u_\mu u_\nu
\biggr\} N +  1/m-{\rm terms} 
\label{LpiN2}
\end{equation}
where the terms not shown explicitly are the ones which by Lorentz
invariance have fixed coefficients, like e.g. $\bar{N} \, D^2/2m \,N$.
Let us make one general remark on the calculation. Whereas the quantities of
interest here, $\sigma(2M_\pi^2)$ and $F_\pi^2\, \bar D^+(0,2M_\pi^2)$, derive
from Lorentz invariant functions, calculations in the heavy baryon formalism
require the choice of a specific kinematical frame. To evaluate the scalar form
factor we choose the Breit frame with $v\cdot (p-p')=0$. Furthermore, to the 
order we are working $\bar D^+(0,2M_\pi^2)$ is given by the spin and isospin
averaged $\pi N$ scattering amplitude in the center--of--mass frame with pion 
energy $v\cdot q = v\cdot q'=M_\pi^2/2m $ and momentum transfer
$(q-q')^2=2M_\pi^2$, disregarding the nucleon pole diagrams.
We have checked that corrections due to this necessary
choice of frame are of order $q^5$ and higher in all cases.
Omitting further calculational details, let us
simply enumerate the results for the various  contributions:
\begin{enumerate}
\item[1)] The terms of the form $M_\pi^4/mF_\pi^2$ and $g_A^4M_\pi^4/mF_\pi^2$
contributing to  $F_\pi^2 \, \bar D^+(0,2M_\pi^2)$ all sum up to zero. For the
latter this seems to contradict the result of ref.\cite{gss},
cf. Eq.(\ref{GSS}). However, only the non--analytic pieces (in the quark mass) 
in the scattering amplitude and the scalar form factor must agree with the 
relativistic calculation and this is  obviously the case here. 
For the finite analytic loop pieces, these do not have to be equal in both
calculations and they can be matched onto each other by appropriate counter 
terms (see ref.\cite{bkkm}). In fact, in all cases where in the relativistic
calculation  one has one--loop functions which have a cut
starting at $t_0= 4m^2$ in the dispersive representation, like for the
nucleon isoscalar electromagnetic and isovector axial radii or the
Goldberger--Treiman discrepancy, one finds a finite piece from the
pertinent one--loop graphs. In the heavy baryon approach, these cuts
are moved to infinity and thus the one--loop graphs have no finite piece.
\item[2)] The terms of the form $g_A^2M_\pi^4/mF_\pi^2 $ give exactly the same
total sum of contributions to $F_\pi^2 \, \bar D^+(0,2M_\pi^2)$ and to
$\sigma(2M_\pi^2)$, namely $3g_A^2 M_\pi^4(\pi-4)/(128\pi^2 m F_\pi^2)$, thus
\beq
\Delta_R^{(g_A^2/m)} =0 \,\,\,.
\eeq
This agrees with the finding in ref.\cite{gss}.
\item[3)] Consider now the loops with exactly one insertion proportional to 
$c_{1,2,3,4}$ (see Fig.1). First, one has to take care of the  renormalization
$F \to  F_\pi$ in the order $q^2$ terms $\sim c_1$ (with $F$ the pion decay
constant in the chiral limit). Both the isoscalar $\pi N$ amplitude at the
Cheng--Dashen point and the scalar form factor of the nucleon at $t=2M_\pi^2$
contain pieces of the type $M_\pi^4 \, \ln M_\pi$, but the resulting
expressions are $identical$ for both
\begin{eqnarray}
& & F_\pi^2 \,\bar D^+(0,2M_\pi^2)^{(c_i-loop)} =
\sigma(2M_\pi^2)^{(c_i-loop)} \nonumber \\
& & = {M_\pi^4 \over 16 \pi^2 F_\pi^2} \biggl\{ 3c_1
\biggl( 8 \ln{M_\pi\over \lambda} + \pi -2 \biggr)
+ c_2\biggl(-2\ln{M_\pi\over \lambda} - {\pi\over4}+{7\over 6}\biggr)- 6c_3
\ln{M_\pi\over \lambda} \biggr\} \nonumber \\ & & \, 
\label{ciloop}
\end{eqnarray}
which is quite an astonishing result\footnote{Note that individually
these contributions to the isoscalar $\pi N$ amplitude and 
$\sigma (2M_\pi^2)$ in  
Eq.(\ref{ciloop}) give numerically about $-11$ MeV 
(for $\lambda \simeq 1$ GeV and $c_{1,2,3}$ taken from \cite{bkmpipin}).
It is therefore not unreasonable to expect the 
remainder $\Delta_R$ coming from the $c_i$-loop graphs to be of similar  
magnitude.} and it implies that 
\begin{equation} 
\Delta_R^{(c_i-loop)} = 0 \quad .
\end{equation}
\item[4)] We have not explicitly calculated all strangeness effects
in $\Delta_R$ but can estimate it from the $K\eta$ loop contribution
to $\sigma(2M_\pi^2)$ at order $M_\pi^4$ \cite{bkmz}
\beq
\sigma(2M_\pi^2)^{(K\eta-loop)} \simeq\frac{5 M_\pi^4}{384 \pi F_\pi^2 M_K}
\simeq 0.4 \, {\rm MeV} \simeq -0.04 \cdot \sigma(2M_\pi^2)^{(c_i-loop)} \,\, .
\eeq
We thus conjecture that the $K\eta$ loop contribution to $\Delta_R$ is
bounded by some fraction of 1 MeV. 
\end{enumerate}

\medskip

\noindent {\bf 3.} To summarize, we have calculated  the remainder at the 
Cheng--Dashen point, $\Delta_R = M_\pi^4
\, C_R$, to order $q^4$ in heavy baryon chiral perturbation theory.
We have proven that $C_R$ has $no$ chiral logarithm and thus it is finite 
in the chiral limit. To this order, only local contact terms
contribute to the remainder. Our estimate
based on the complete $q^4$ CHPT calculation with the low energy constants of
the pertinent counter terms saturated via resonance exchange is
\begin{equation} \Delta_R \approx 2 \, {\rm MeV} \,\,\, ,
\end{equation}
which we consider a conservative upper bound. As conjectured in ref.\cite{gls}
the remainder $\Delta_R$ indeed does not play any role in the extraction of the
$\sigma$--term from the $\pi N$ data considering the present status of accuracy
of these data in the threshold region.

\bigskip  

\section*{Acknowledgements}
This work grew out of a discussion during the workshop ``The Standard
Model at Low Energies'' at the ECT*, Trento, May 1996.
We are grateful to J\"urg Gasser for some valuable comments and
encouragement.

\bigskip


\section*{Figure}
\begin{enumerate}
\item[Fig.1] Loop diagrams which lead to the result given in 
Eq.(\ref{ciloop}). 
The circle--cross denotes an insertion from ${\cal L}_{\pi N}^{(2)}$
proportional to $c_{1,2,3}$. Full, broken and wavy lines represent nucleons,
pions and the external scalar source, respectively. Subsets of diagrams which
add up to zero are not shown.  
\end{enumerate}

$\,$

\vskip 1cm

\begin{figure}[bht]
\centerline{
\epsfysize=4.2in
\epsffile{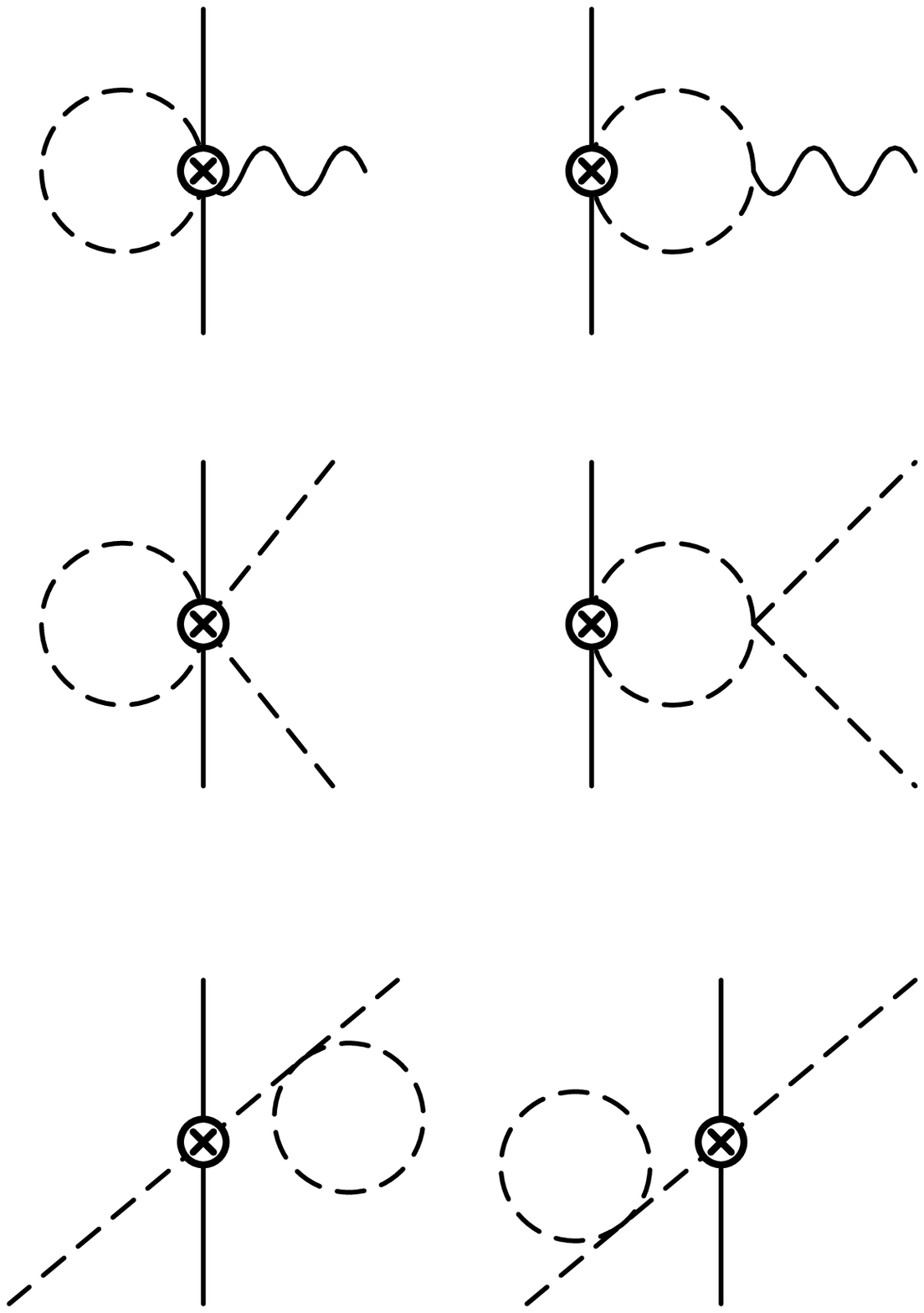}
}

\vskip 1.5cm

\centerline{\Large Figure 1}
\end{figure}

\end{document}